\documentclass[12pt,preprint]{aastex}

\slugcomment{\small Published online in {\it Science Express} (26 Feb. 2004) \\
To appear in print in $Science$ (2004)}
\shorttitle{Discovery of a dust disk around AU Mic}
\shortauthors{Kalas, Liu and Matthews.}

\begin{document}

\title{Discovery of a large dust disk around \\
 the nearby star  AU Microscopium}

\author{Paul Kalas\altaffilmark{1,2}, Michael C. Liu\altaffilmark{3}, Brenda C. Matthews\altaffilmark{1}}
\affil{}
\altaffiltext{1}{Astronomy Department and Radio Astronomy Laboratory, 
601 Campbell Hall, Berkeley, CA 94720}
\altaffiltext{2}{National Science Foundation Center for Adaptive Optics, University of California, Santa Cruz, CA 95064}
\altaffiltext{3}{Institute for Astronomy, 2680 Woodlawn Dr., Honolulu, HI 96822}

\begin{abstract}
We present the discovery of a circumstellar dust disk 
surrounding AU Microscopium (AU Mic, GJ 803, HD 197481). 
This young M star at 10 parsec has the 
same age and origin as $\beta$ Pictoris, 
another nearby star surrounded by a dust disk.
The AU Mic disk is detected between 50 AU and 210 AU radius, 
a region where dust lifetimes exceed the present stellar age.  
Thus, AU Mic is the nearest star where we 
directly observe the solid material required for
planet formation.  Since 85\% of stars are M-type,
the AU Mic disk provides new clues on how the
majority of planetary systems might form and evolve.
\end{abstract}


\section{Introduction}

About 15\% of nearby main-sequence stars exhibit an excess of far-infrared radiation
that points to the existence of circumstellar dust grains ({\it1}).  
Dust grains have short lifetimes, and their continued presence implies a source
of replenishment.  The solar system has a disk-like distribution of dust that is
continually replenished by the sublimation of comets and collisions between asteroids.
We therefore infer that similar populations of undetected parent bodies produce
dusty debris disks around infrared excess stars.  
Moreover, evidence for planets can be found by matching density variations in
debris disks to theoretical models of how planets gravitationally perturb these disks
({\it2}, {\it3}, {\it4}).
In effect, circumstellar debris disks
are a signpost for the existence of extrasolar planetary systems.  

Direct images of debris disks are rare.  Starlight reflecting off optically thin debris disks
is detected at optical and near-infrared wavelengths in only three cases - 
$\beta$ Pictoris, HR 4796A and HD 141569 ({\it5}, {\it6}, 
{\it7}).  
In three more cases - Vega, Fomalhaut, and $\epsilon$ Eridani - 
debris disk structure is seen only at thermal infrared wavelengths 
({\it8}, {\it9}).  
$\beta$ Pic, HR 4796A and HD 141569 are relatively young ($<$20 Myr)
main sequence stars that have the largest disk masses, and
hence represent the more detectable of the debris disk systems.

$\beta$ Pic has Galactic space motions in common with two M stars, 
AU Mic and AT Mic, that have ages $\sim$20 Myr ({\it10}).  
These stars and $\beta$ Pic may be
co-eval sister stars that have separated in space over time due to the
small differences in their space motions.  A total
of 17 stars may be members of this group with age 8$-$20 Myr ({\it11}).
AU Mic has significant infrared excess at 60 $\mu$m 
({\it12}, {\it13}) and recent
sub-millimeter data reveal the
the presence of cold (40 K) dust, a total dust mass roughly three times smaller
than that of $\beta$ Pic (Table 1), an 
absence of molecular gas, and a lack of grains within 17 AU
of the star ({\it14}).

\section{Observations}

We endeavored to directly image circumstellar dust around AU Mic with 
an optical stellar coronagraph at the University of Hawaii 2.2-m
telescope on Mauna Kea, Hawaii ({\it15}).  The stellar coronagraph (Fig. S1) produces 
artificial eclipses of stars by blocking the light at the focal plane with
a circular opaque mask suspended by four wires.  Instrumental
diffracted light is blocked by a Lyot mask in the pupil plane.  The net
result is significantly enhanced contrast in the regions surrounding
bright stars.  The imaging camera behind the coronagraph is a Tek
2048$\times$2048 CCD with a scale of 0.41$\arcsec$/pixel.  All our
data were obtained through a standard broadband $R$ filter ($\lambda_c$=647 nm,
$\Delta\lambda$ = 125 nm).

We acquired data on 14 and 15 October, 2003, with 6.5$\arcsec$
and 9.5$\arcsec$ diameter occulting spots, respectively.  We obtained five
240-s and three 300-s images of AU Mic on the first and second nights,
respectively.   In addition to
AU Mic, we observed five other bright stars to
check for spurious features such as diffraction spikes and internal reflections (Table S1).
A disk-like reflection nebulosity surrounding AU Mic 
was detected in raw data during both nights of observation and
does not match the position angle (PA, measured east from north), 
width, or morphology of instrumental diffraction
spikes.  The image quality as measured by the full-width at half-maximum (FWHM)
of field stars was $\sim$1.1$\arcsec$.

Data reduction followed the standard steps of bias subtraction, flat-fielding
and sky-subtraction.  We then subtracted the stellar
point spread function (PSF) to remove
excess stellar light from around the occulting spot.  We used the real PSF's
from other stars observed throughout each night, as well as artificial PSF's.
Artificial PSF subtraction is effective for AU Mic because the circumstellar disk is close to
edge-on.  We extracted the stellar PSF for each image of AU Mic
by sampling the image radially in
a direction perpendicular to the PA of the disk.
We then fit a polynomial to the data and generated an 
artificial PSF that is a figure of rotation of the polynomial.  The PSF's were then scaled
and registered to each data frame such that subtraction minimized 
the residual light in directions perpendicular to the disk beyond the edge of the
occulting spot.  In general, the different PSF subtractions produce comparable
results, with minor differences appearing a few pixels beyond the edge
of the occulting spot.  
To evaluate the uncertainties in our final image, we measured disk properties
in four different data sets that represent two nights of observation, and each with
two different PSF subtraction techniques.  The data for $\beta$ Pic
from ({\it16}) that are discussed below were obtained with the same
telescope, coronagraph, and filter, and analyzed using similar techniques.

\section{Circumstellar Dust Morphology}

The reflection nebulosity around AU Mic is consistent
with a circumstellar disk seen at a near edge-on viewing geometry (Fig. 1, Fig. S2).   
We detect the disk as far as $\sim$21$\arcsec$  (210 AU) from the star ({\it17}).
This sensitivity-limited value is a lower limit to the true disk outer radius.  
The inner radius of the detected
disk is 5$\arcsec$ (50 AU) and is mainly
limited by the radius of the occulting spot and artifacts of the PSF subtraction.  
The position angles of the two disk midplanes differ by about 6$\pm$3$\degr$;  PA = 
124$\pm$2$\degr$ for the SE side and PA = 310$\degr\pm$1$\degr$ for the NW side.
A similar, 1.0$\degr-$2.5$\degr$
offset,  called the ``wing-tilt asymmetry'',  was measured for the $\beta$ Pic midplanes ({\it16}).  
A symmetric disk can
appear to have a wing-tilt when the disk axis is tilted to the line of sight
and the scattering phase function is non-isotropic.  
A model-dependent relationship between the observed wing-tilt and
the intrinsic disk inclination ({\it16}) 
suggests that AU Mic disk is inclined 7$\degr-$20$\degr$
from edge-on.  
On the other hand, the sharp midplane morphology is 
consistent with model disks that have 
inclinations no greater than $\sim$5$\degr$ from edge-on ({\it18}, Fig. S3).   
Until higher resolution data are obtained and analyzed,  
we adopt a disk inclination of $\sim$5$\degr$ from edge-on.

Power-law fits to the disk midplanes between 6$\arcsec$ (60 AU) and 16$\arcsec$ (120 AU)
radius give indices of -3.6 and -3.9 for the NW and SE extensions,
respectively (Fig. 2).  These indices are similar to the power-law
fits for the NE and SW disk extensions of $\beta$ Pic (Table 1)({\it16}).
However, the midplane profiles for $\beta$ Pic become less steep 
inward of 100 AU.  No such turnover is seen for the AU Mic
radial profiles.  
The NW midplane of AU Mic also shows a significant enhancement
in surface brightness $\sim$9$\arcsec$ radius from the star (Fig. 1, Fig. 2).  This
could be due to a background source, but further tests using color,
polarization and proper motion information should be evaluated before excluding
a physical connection to AU Mic.  

\section{Discussion}

The existence of morphologically similar dust disks around AU Mic and $\beta$ Pic
supports the hypothesis that these are sister stars born at the same
time and location.  However, the two disks are not twins.  
The total mass of dust estimated
from the spectral energy distributions is 3.3 times greater for
$\beta$ Pic relative to AU Mic (Table 1).  The relative brightnesses
of the two disks in optical data are consistent with this result.
To make the comparison, we imagine placing the $\beta$ Pic
dust disk around AU Mic.  
In Figure 2, we include the midplane surface brightness
profile for $\beta$ Pic using data from ({\it16}) that
is now scaled by factors that
account for the AU Mic heliocentric distance and
stellar luminosity.  We find that if the
disk of $\beta$ Pic surrounded AU Mic it would be about
1.5 mag arcsec$^{-2}$ brighter than what we measure for the AU Mic disk (Fig. 2).
This corresponds to a factor of four greater scattering
cross section of $\beta$ Pic grains relative to AU Mic grains.  
If we assume that the two disks have 
exactly the same structure, grain properties, and viewing geometry
then the AU Mic disk requires a dust mass that is four times smaller
than that of $\beta$ Pic.  Future observations of disk properties such as the 
inclination of AU Mic will elucidate the validity of these assumptions, 
but this result is consistent with the infrared dust luminosity.  

The underlying grain properties are also likely to differ due to the
weak radiation environment of an M star relative to an A star.
AU Mic is 3.6 times less massive than $\beta$ Pic, and 87 times less
luminous (Table 1).  
For the AU Mic disk, the collision timescale at 100 AU radius is 0.2$-$1.8 Myr assuming a 
dust optical depth of $\tau\sim$10$^{-3} -10^{-4}$, respectively (Fig. S4).  At
200 AU, near the outer boundary of the detected disk, the collision timescale is
0.5$-$5.0 Myr.  Given an age of 8$-$20 Myr for AU Mic, most disk particles have
undergone at least one collision.  However,
as objects are shattered into smaller pieces, the radiation
pressure force around AU Mic is too weak to remove the fragments ({\it19}).  They 
can be removed by the system either by joining together to form larger
objects, or by spiraling into the star by Poynting-Robertson (PR) drag.  The
PR timescales at 100 AU are 0.2$-$1.8 Gyr for 1$-$10 $\mu$m particles, respectively $-$
many times longer than the age of the system ({\it1}).  For $\beta$ Pic, on the 
other hand, grains a few microns and smaller are quickly ejected by 
radiation pressure and the disk mass diminishes over time ({\it20}).  The AU Mic
disk should preserve a larger population of sub-micron sized
grains, and the mass of solid objects observed today should approximate 
the primordial disk mass.  In other words, most of the disk 
seen in our optical scattered light image may consist of primordial solid
material.  

Within $\sim$50 AU of the star, the timescales for grain removal by collisions and PR drag become 
significantly shorter than the stellar age.  Primordial dust at the inner
limit of our images (Fig. 1, 2) has mostly vanished,
and the grains observed here, as well as those discovered as close 
as 17 AU from the star ({\it14}), must be continually 
replenished by the collisional erosion
of much larger objects such as comets and asteroids.  
The existence of planetesimals in this region
lends plausibility to the argument that the same objects will form planets by
accretion.  Given that AU Mic is only $\sim$10 Myr old, we may be able
to observe planets that are still in the process of accreting mass, or at least
discern disk structure that is sculpted by planet-mass bodies.  Because
AU Mic is closer to the Sun than $\beta$ Pic, the 2$-$30 AU zone where
terrestrial and gas giant planets might form can be resolved by current and 
future instrumentation (Fig. S5).  Planets around AU Mic may also be detected by
indirect methods.
The low stellar mass means that the star will 
display a significant astrometric reflex
motion (2 milli-arcsec for a Jupiter-analog).  
The near edge-on orientation favors planet detection by transits
of the stellar photosphere.  Finally, if a planet is detected by radial velocity techniques,
then the near edge-on orientation gives the planet mass by constraining the sin($i$)
ambiguity intrinsic to these measurements.

\newpage
\noindent{\bf References and Notes}

\noindent1. 
D. E. Backman, F. Paresce, in {\it Protostars and Protoplanets III}, 
E. H. Levy and J. I. Lunine, Eds. (University of Arizona Press, Tucson, 1993), pp. 1253-1304.

\noindent2.
F. Roques, H. Scholl, B. Sicardy, B. A. Smith, $Icarus$ {\bf 108}, 37 (1994).

\noindent3.
J. -C. Liou, H. A. Zook, {\it Astron. J.} {\bf 118}, 580 (1999).

\noindent4.
L. M. Ozernoy, N. N. Gorkavyi, J. C. Mather, T. A. Taidakova {\it Astrophys. J.} {\bf 537}, L147 (2000).

\noindent5. 
B. A. Smith, R. J.Terrile, $Science$ {\bf 226}, 1421 (1984).

\noindent6.
G. Schneider et al., {\it Astrophys. J.} {\bf 513}, L127 (1999).

\noindent7.
M. D. Silverstone et al., {\it Bull. Am. Astron. Soc} {\bf 30}, 1363 (1998).

\noindent8.
W. S. Holland et al., $Nature$ {\bf 392}, 788 (1998).

\noindent9.
J. S. Greaves et al., {\it Astrophys. J.} {\bf 506}, L133 (1998).

\noindent10.
D. Barrado y Navascues, J. R. Stauffer, I. Song, J-. P. Caillault,
{\it Astrophys. J.} {\bf 520}, L123 (1999).

\noindent11.
B. Zuckerman, I. Song, M. S. Bessell, R. A. Webb,  {\it Astrophys. J.} {\bf 562}, L87 (2001).

\noindent12.
V. Tsikoudi, {\it Astron. J.} {\bf 95}, 1797 (1988).

\noindent13.
I. Song, A. J. Weinberger, E. E. Becklin, B. Zuckerman, C. Chen, 
{\it Astron. J.} {\bf 124}, 514 (2002).

\noindent14.
M. C. Liu, B. C. Matthews, J. P. Williams, P. G. Kalas, {\it Astrophys. J.}, in press (2004).

\noindent 15. 
Materials and methods are available as supporting material on {\it Science Online}.

\noindent16.
P. Kalas, D. Jewitt, {\it Astron. J.} {\bf 110}, 794 (1995).

\noindent17. To increase the signal-to-noise of the data shown in Fig. 1, we 
binned the data 3$\times$3 pixels and then smoothed by a Gaussian function
with $\sigma$=0.5 pixel.   This smoothed
image was used to find the maximum outer extent of the disk.  All other measurements
were made using the unbinned and unsmoothed image shown in Fig. 1.

\noindent18. 
P. Kalas, D. Jewitt,  {\it Astron. J.} {\bf 111}, 1347 (1996).

\noindent19.
R. Saija, et al. {\it Mon. Not. R. Astron. Soc.} {\bf 341}, 1239 (2003).

\noindent20.
P. Artymowicz, {\it Astrophys. J.} {\bf 335}, L79 (1988).

\noindent21.
M. S. Bessel,  F. Castelli,  B. Plez, {\it Astron. Astrophys.} {\bf 333}, 231(1998).

\noindent22.
F. Crifo, A. Vidal-Madjar, R. Lallement, R. Ferlet, M. Gerbaldi,  {\it Astron. Astrophys.}
 {\bf 333}, L29 (1997).

\noindent23.
J. D. Larwood, P. G. Kalas, {\it Mon. Not. R. Astron. Soc.} {\bf 323}, 402 (2001).

\noindent24.
W. R. F. Dent, H. J. Walker, W. S. Holland, J. S. Greaves, 
{\it Mon. Not. R. Astron. Soc.} {\bf 314}, 702 (2000).

\noindent25.
This work has been supported by the NASA Origins Program under grant NAG5-11769,
and the NSF Center for Adaptive Optics, managed 
by the University of California at Santa Cruz under cooperative 
agreement No. AST-9876783.    BCM acknowledges support
from NSF grant \#0228963.  MCL acknowledges support from
a Hubble Postdoctoral Fellowship (NASA Grant HST-HF-01152.01).
The authors acknowledge the insightful contributions 
of two anonymous referees.

\noindent {\bf Supporting Online Material}\\
www.sciencemag.org\\
Materials and Methods\\
Figs. S1, S2, S3, S4 and S5\\
Table S1\\

\newpage
\begin{figure}
\epsscale{0.8}
\plotone{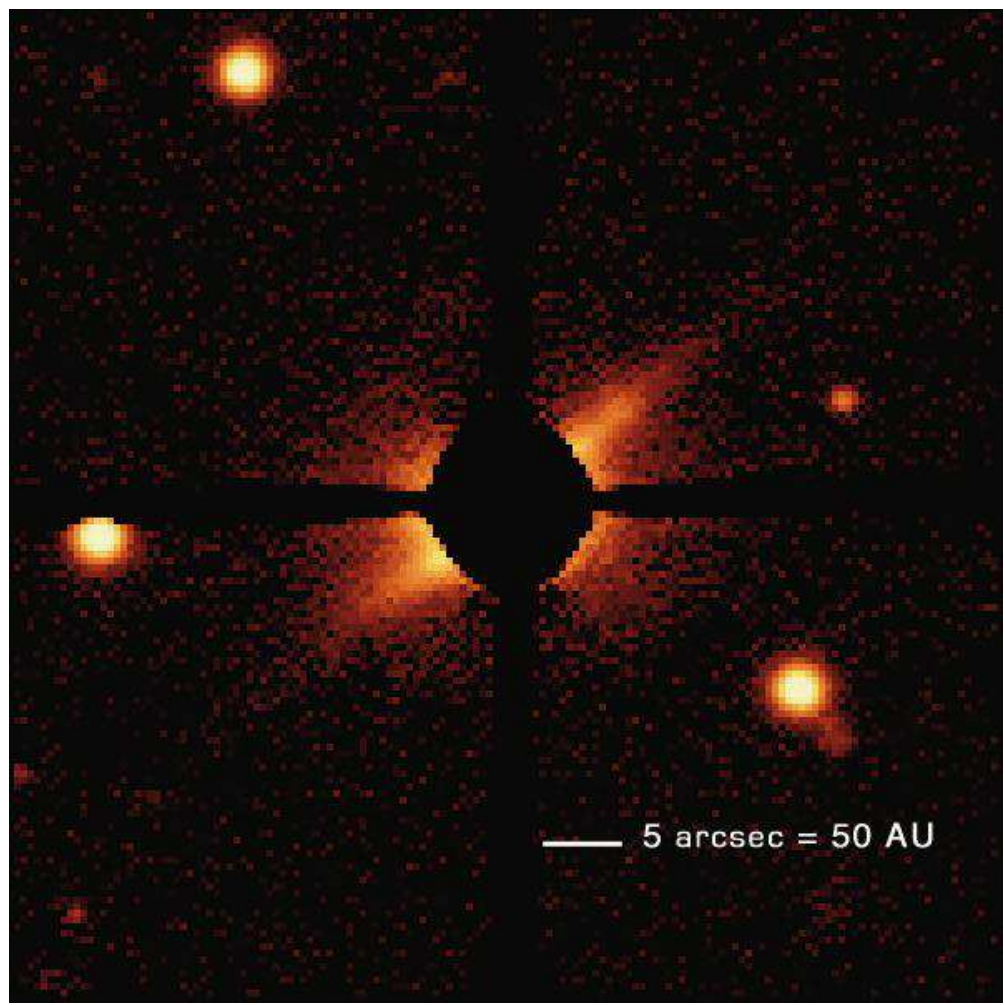}
\caption{
The disk surrounding AU Mic seen in optical scattered light.  North is up, east is left,
and each side of this false-color image corresponds to 60$\arcsec$.  The central dark region is produced by
the 9.5$\arcsec$ diameter focal plane occulting spot which is suspended by four wires and completely
masks our direct view of the star.  This image represents 900 seconds total integration in the $R$ band
and each pixel corresponds to 4 AU at the distance to AU Mic.  
Residual light evident near the occulting spot edge in the NE-SW direction is attributed
to asymmetries in the point-spread function caused by instrumental scattering and atmospheric
seeing.
 \label{fig1}}
\end{figure}

\begin{figure}
\epsscale{0.8}
\plotone{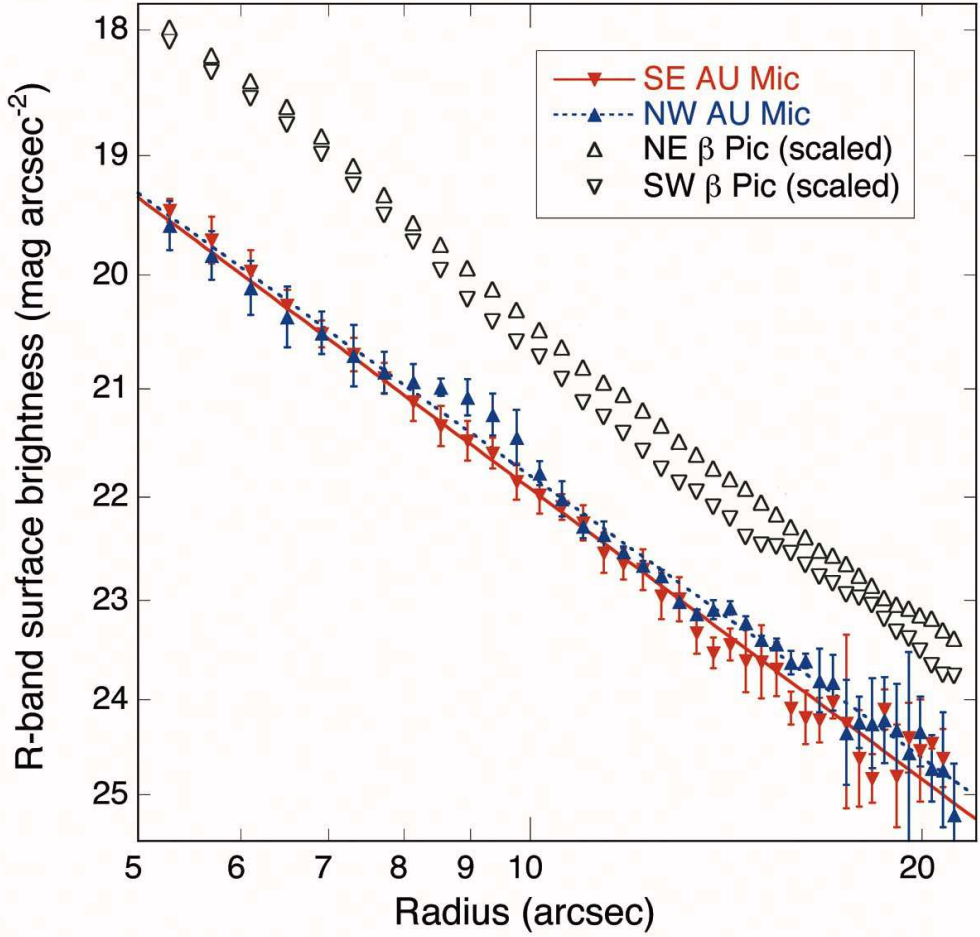}
\caption{\small
Midplane surface brightness as a function of radius.  The midplane was
sampled between 5$\arcsec$ and 21$\arcsec$ radius along a strip 1.2$\arcsec$ wide.
We show the mean value from two nights of data with two different PSF
subtraction techniques.  The error bars represent one standard deviation of a single measurement.  
We fit the data between 6$\arcsec$ and 16$\arcsec$ radius 
with power laws that give indices -3.6 and -3.9 for the NW and SE
midplanes of AU Mic, respectively.   The radial  profile for the NW midplane
has a significant brightness enhancement at $\sim$9$\arcsec$ radius that is either 
intrinsic to the disk or a background object. We also plot
the surface brightness of the $\beta$ Pic disk from
({\it16}), but with the surface brightness uniformly 3.0 mag arcsec$^{-2}$
fainter to simulate the existence of $\beta$ Pic's disk around AU Mic at 9.9 pc.  
This scaling takes into account the fact that the absolute $R$-band magnitude
of AU Mic is 5.6 mag fainter than $\beta$ Pic (Table 1), and at a constant angular
radius the $\beta$ Pic disk is roughly a factor of 
($d_{AU Mic} / d_{\beta Pic})^{-3.6}$ = (9.9 / 19.3)$^{-3.6}$ =11.3 times brighter
(i.e. 2.6 mag brighter;  see Eqn. 4 in {\it18}).  
\label{fig2} }
\end{figure}

\clearpage 

\begin{deluxetable}{lllll}
\tabletypesize{\small}
\tablecaption{Star (rows 1-8) and disk (rows 9-13) properties for AU Mic and $\beta$ Pic.
The stellar parameters for AU Mic are derived from 
data given by ({\it10}, {\it13}, {\it21}).
For $\beta$ Pic's stellar parameters, we use ({\it22}) and
references therein.  The $\beta$ Pic disk $R$-band surface brightness (SB) at
6$\arcsec$ radius is given in ({\it16}), while its maximum extent
is given in ({\it23}).
\label{tbl-1}}
\tablewidth{0pt}
\tablehead{
\colhead{} & \colhead{AU Mic} & \colhead{$\beta$ Pic}
}
\startdata
Spectral Type 						&M1Ve 		& A5V \\
Mass (M$_\odot$)					&0.5			&1.8\\
T$_{eff}$ (K) 						&3500 		& 8200\\
Luminosity (L$_\odot$)				&0.1			&8.7\\
Distance (pc)  						&9.9			& 19.3 \\
$V$ (mag)						&8.8		& 3.9 \\
M$_V$ (mag)						&8.8		& 2.4 \\
$V-R$							&0.88		& 0.08\\
Disk SB (6$\arcsec$)				&20.1$\pm$0.3	&15.4$\pm$0.3\\
SB fall-off\tablenotemark{a}			&-3.6 to -3.9	&-3.8 to -4.1 \\
Max. Radius (AU)					&210		&1835  \\
$\tau$=$L_{disk}$/$L_{bol}$\tablenotemark{b}	&6.1$\times$10$^{-4}$ & 3$\times$10$^{-3}$ \\
Total Dust Mass (g)\tablenotemark{c}	&6.6$\times$10$^{25}$& 2.2$\times$10$^{26}$ \\

\enddata
\tablenotetext{a}{Value of exponent for a power-law fit to disk $R$-band surface
brightness fall-off between 6-16$\arcsec$ radius.  The shallower surface
brightness profiles correspond to the NW and NE brightness profiles of
AU Mic and $\beta$ Pic, respectively.  The $\beta$ Pic values
are taken from ({\it16}).
}
\tablenotetext{b}{The fractional dust luminosity, assuming an optically thin disk,
determined by taking the ratio of excess infrared luminosity to stellar bolometric luminosity.
The values are obtained from  ({\it14}) and ({\it1}) for
AU Mic and $\beta$ Pic, respectively.
}
\tablenotetext{c}{Dust mass from model fits to the spectral energy
distributions taken from ({\it14}) for AU Mic, and ({\it24}) for
$\beta$ Pic.
}
\end{deluxetable}
\clearpage

\section*{SUPPORTING ONLINE MATERIAL \\
Materials and Methods}

\begin{figure}[th]
\epsscale{0.7}
\plotone{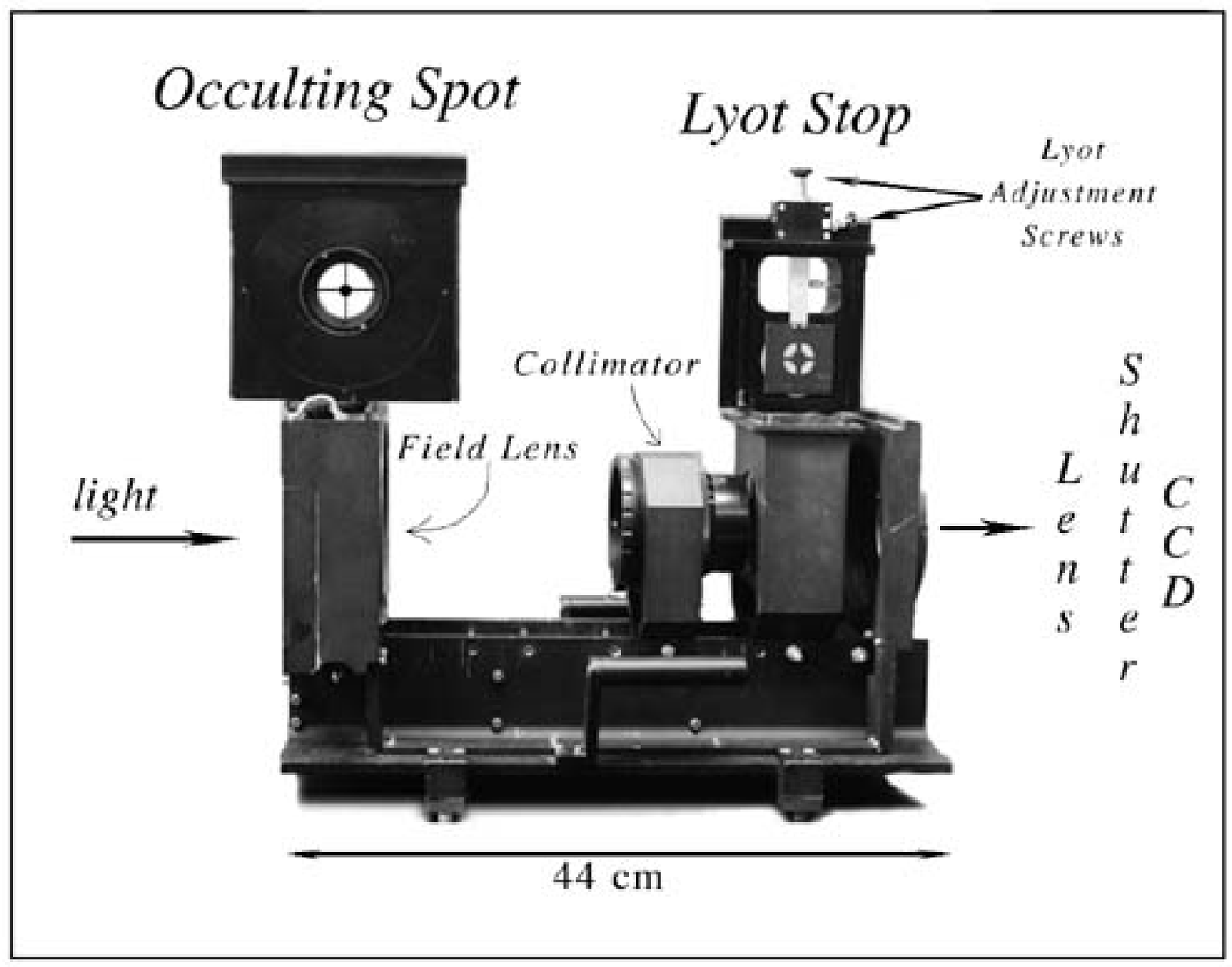}
\begin{flushleft}
{{\bf Fig. S1: }
The optical stellar coronagraph used for these observations. The main function of a stellar coronagraph is to block light from a bright star in order to detect faint, nearby objects. The occulting spot is placed at the focal plane of the telescope and prevents light from striking any optical elements further down in the optical path. Without the occulting spot, light from the star would saturate the CCD, and the optical elements would fill the background with scattered light, as well as produce spurious reflections. The optical elements in the coronagraph are used to create an image of the pupil plane, essentially an image of the telescope mirrors, and support structures. The Lyot stop is appropriately shaped to block this image of the telescope diffraction pattern - stars imaged with a coronagraph do not have diffraction spikes. The removal of the diffraction pattern is a major advantage when trying to image faint sources near the star.
\label{fig1}}
\end{flushleft}
\end{figure}

\begin{figure}
\epsscale{0.6}
\plotone{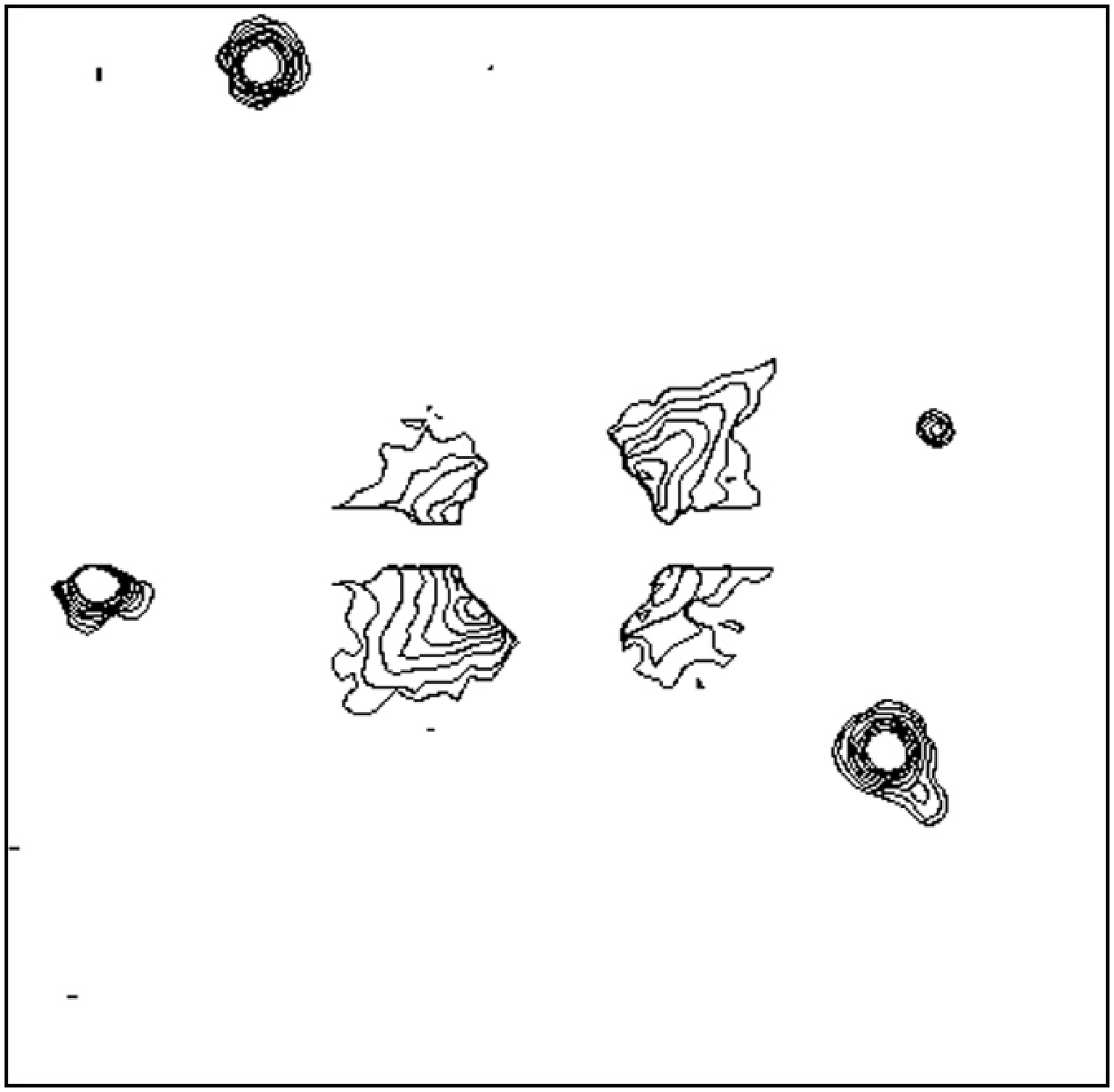}
\begin{flushleft}
{{\bf Fig. S2: }
R-band surface brightness map corresponding to Fig. 1.  Isophotes are plotted at 0.5 mag arcsec$^{-2}$ 
intervals from 23.5 to 20.0 mag arcsec$^{-2}$ .  
As with Fig. 1, north is up, east is left, and the sides of the box represent 60$\arcsec$.
 \label{fig1}}
 \end{flushleft}
\end{figure}

\begin{figure}
\epsscale{1.0}
\plotone{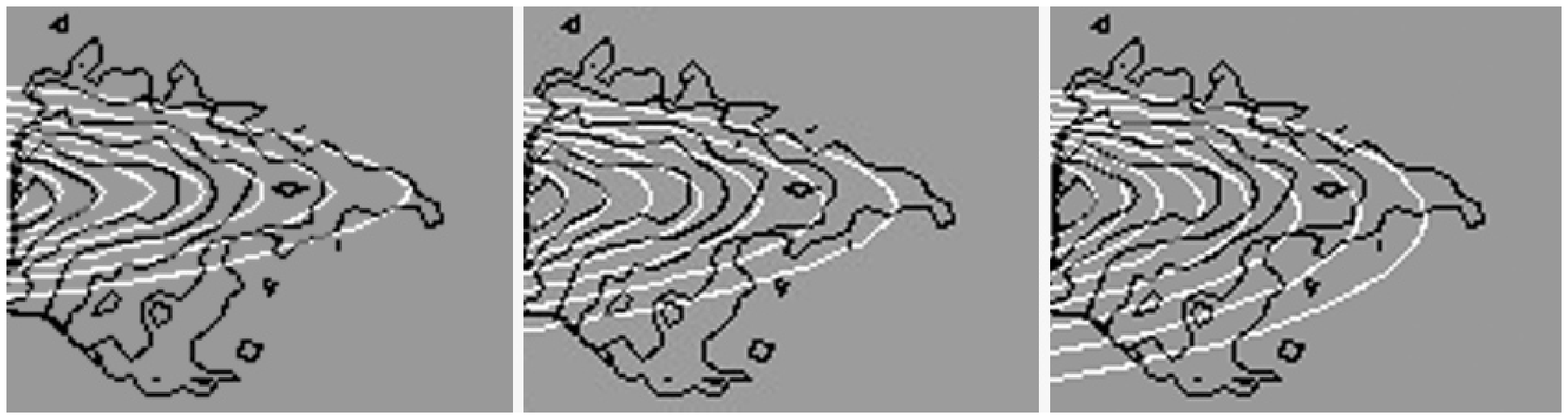}
\begin{flushleft}
{{\bf Fig. S3: } Comparison of the AU Mic disk to model disks at different inclinations to the line of sight.  We rotate the surface brightness map of AU Mic such that the midplane of the NW disk extension is oriented horizontally (black contours).  The three panels above show a model disk (white contours) that is inclined edge-on (left panel), 5$\degr$ from edge-on (middle panel), and 10$\degr$
from edge-on (right panel).  The model used is identical to the $\beta$ Pic model described in Section 4.1 of (1).  It is normalized to the AU Mic surface brightness at 6$\arcsec$ radius (Fig. 2), and convolved with a Gaussian to simulate the
seeing conditions during the observations.   
In the dust disk model, the midplane volume number density distribution decreases with radius as a power-law with index -2.6.  All model parameters are kept the same in the three models above, except for the disk inclination to the line of sight.  Our main conclusion is that the AU Mic disk strongly resembles the $\beta$ Pic disk in its morphology and disk inclination.  The model that is 10$\degr$ away from edge-on (right panel) demonstrates isophotes that are significantly more rounded at the midplane than what we observe for AU Mic.  Thus, given the limited $\sim$1.1$\arcsec$ resolution of our observations, we constrain the disk inclination to no greater than $\sim$5$\degr$ from edge-on.   
 \label{fig1}}
 \end{flushleft}
\end{figure}

\begin{figure}
\epsscale{0.6}
\plotone{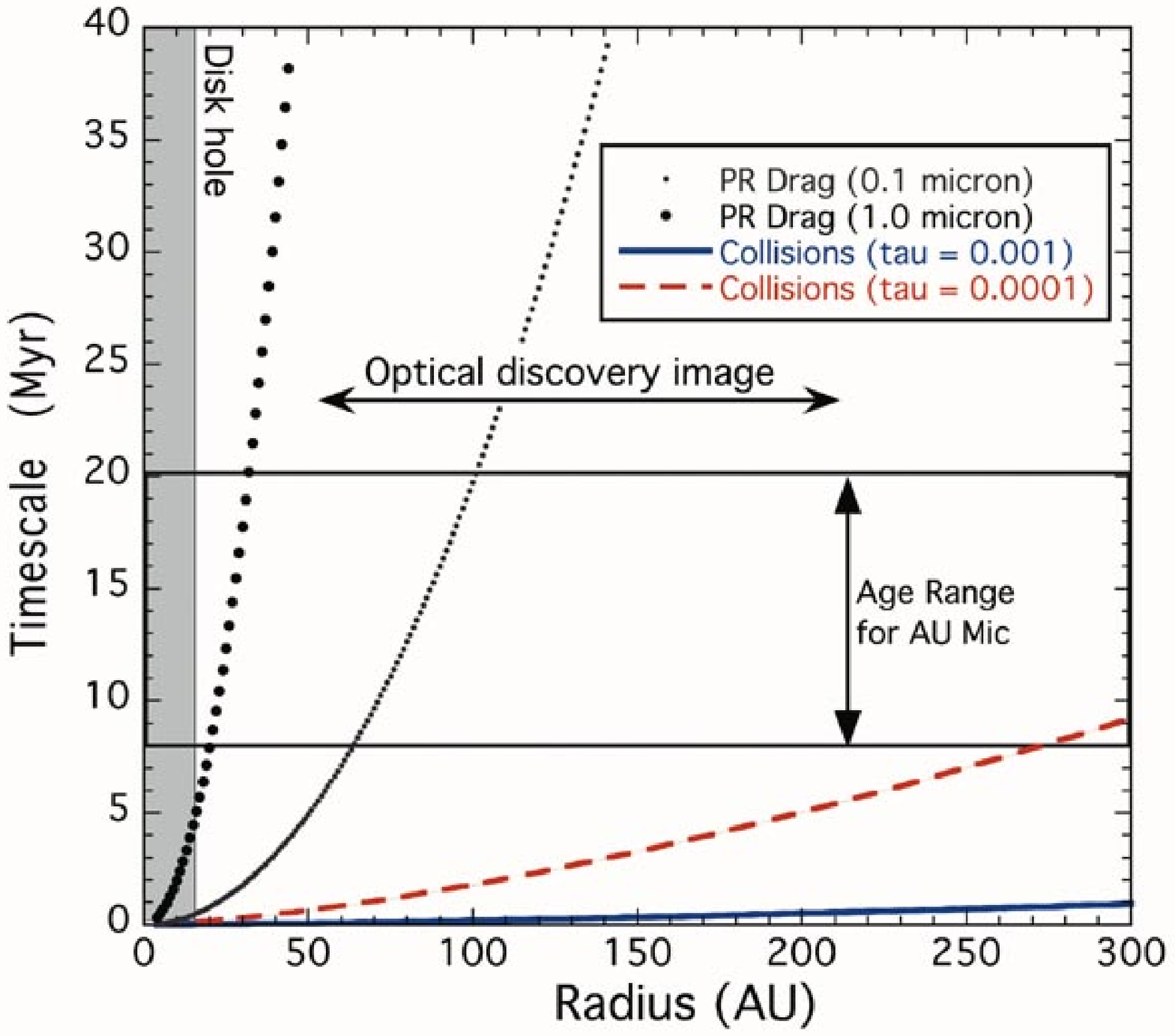}
\begin{flushleft}
{{\bf Fig. S4: }
\small
Dust lifetimes as a function of radius.  After gas in a circumstellar disk has dispersed, and the system becomes optically thin, Poynting-Robertson (PR) drag causes grains to spiral into the star.  Using the equation for PR drag timescales given by (2), we plot the PR drag timescales for 0.1 $\micron$ and 1.0 $\micron$ grains.  At radii $<$67 AU, the 0.1 $\micron$ grains have spiraled into the star by 8 Myr.  The 1.0 $\micron$ grains, on the other hand, take much longer to spiral into the star if they are initially located at 67 AU radius.  In fact, only if they were within $\sim$20 AU radius would 1.0 $\micron$ grains have enough time (8 Myr) to spiral into the star.  In the region between 20-67 AU we would expect to find a lack of 0.1 $\micron$ grains.  Beyond 67 AU, the 0.1 $\micron$ grains have not had enough time to spiral into the star.  The disk scattered light may appear bluer at these larger radii due to Raleigh scattering. 
	The collision timescales are also shown (2).  We assume two values for optical depth, $\tau$, that span a likely range of values inferred from the observed spectral energy distribution (Table 1).  At 100 AU radius, and for $\tau$=0.0001, the
collision timescale is 1.8 Myr.  If the age is 8 Myr, then grains are likely to have had at least one collision.  Collisions tend to be destructive, increasing the population of submicron grains.  Thus most of the grains seen in the discovery image (Fig. 1) are pieces of larger objects, and this is why such disks are called "debris disks".  Unlike very luminous stars such as $\beta$ Pic and Vega, the submicron grains orbiting AU Mic will not be ejected outward by radiation pressure.  Therefore, the disk around AU Mic should retain a larger fraction of its primordial material.  Beyond $\sim$200 AU radius we expect the disk to be collisionally un-evolved.  The grains here may represent the pristine material in interstellar clouds that are the building blocks for larger particles that join to form planetesimals such as comets and asteroids. 
Also shown above is a gray region that indicates the inferred radius of a hole in the disk (3).  The force of PR drag discussed above should cause grains to spiral into this region from the outer parts of the disk.  The fact that very warm material is not evident at infrared wavelengths indicates that grains are prevented from approaching closer than $\sim$17 AU radius.  The existence of the hole hints at the possibility of a planet-mass object orbiting the star because planet-grain encounters are one mechanism for efficient grain removal. \\
\label{fig1}}
\end{flushleft}
\end{figure}

\begin{figure}
\epsscale{1.0}
\plotone{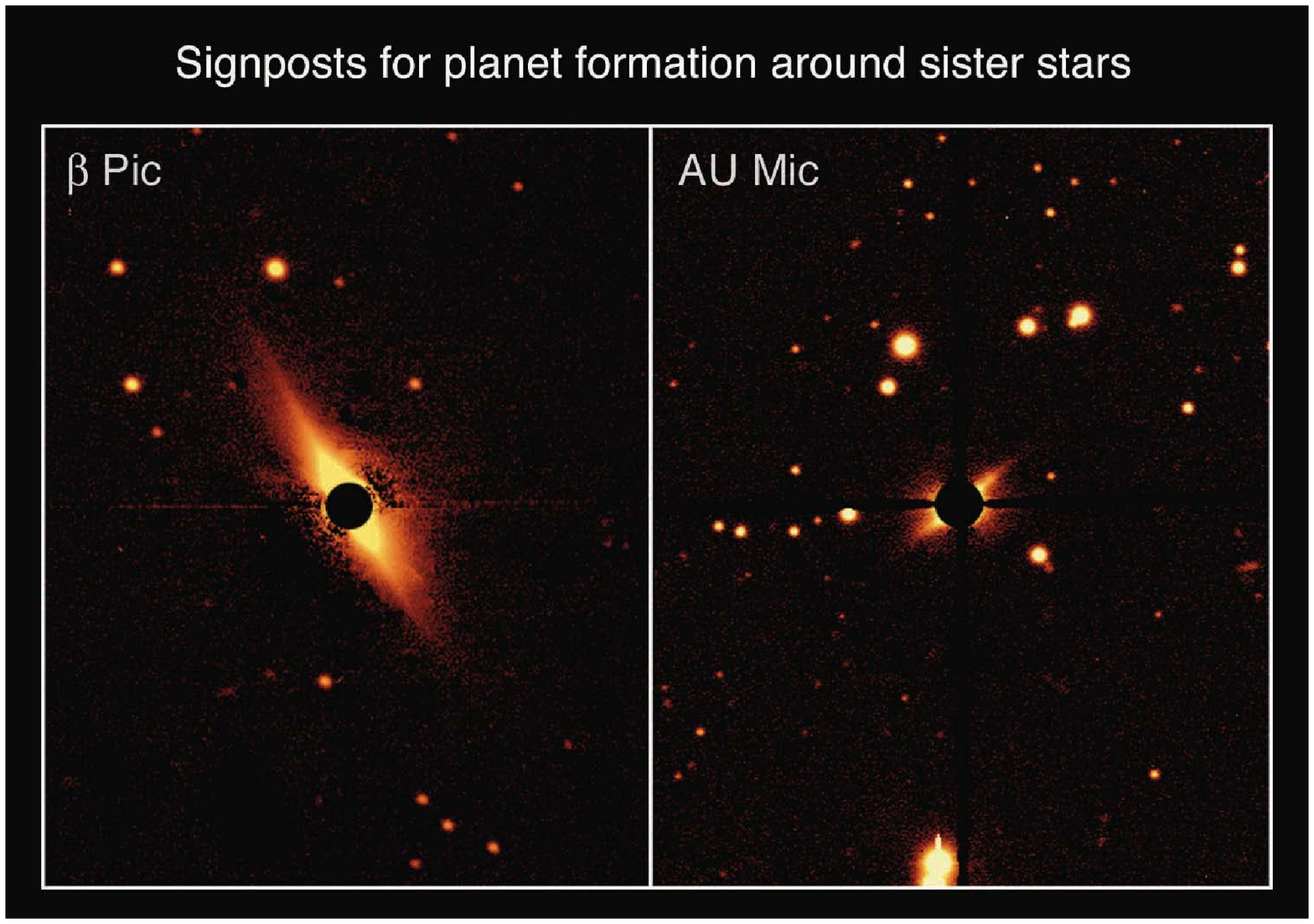}
\begin{flushleft}
{{\bf Fig. S5: }
Beta Pic on the left (4) was imaged with the same telescope, filter, and instrument as AU Mic on the right (this paper).  The total integration times are nearly equal, providing roughly the same sensitivity. The $\beta$ Pic disk appears brighter because it surrounds a brighter star, and it contains more dust grains than the AU Mic disk.   These false-color images are also shown to the same angular scale - the circular black masks have diameter 10$\arcsec$. Because $\beta$ Pic is about twice as distant as AU Mic is, 10$\arcsec$ corresponds to 193 astronomical units at $\beta$ Pic, and only 99 astronomical units at AU Mic.  Therefore, in the case of AU Mic, the dust disk observed near the boundary with the black mask approximately corresponds to the Kuiper Belt in our solar system.  The proximity and youth of AU Mic make it an ideal target to directly study the possible planet-forming environment hidden behind the mask used in our observations.  Ground-based telescopes with adaptive optics and the Hubble Space Telescope are well suited for examining this region.  Together, AU Mic and $\beta$ Pic present an exciting opportunity to understand how two planetesimal systems born at the same time will evolve differently.  
 \label{fig1}}
 \end{flushleft}
\end{figure}

\clearpage

\setcounter{table}{18}
\def\thetable{\Alph{table}}

\begin{deluxetable}{lllll}
\tabletypesize{\small}
\tablecaption{1 : PSF reference stars used in the analysis of the AU Mic data}
\tablewidth{0pt}
\tablehead{
\colhead{Name} & \colhead{RA (J2000)} &  \colhead{DEC (J2000)} &\colhead{V(mag)} & \colhead{SpT} 
}
\startdata
HD 17925	&	02 52 32	&-12 46 10&		6.0	&	K1V\\
HD 22049	&	03 32 56	&-09 27 30	&	3.7	&	K2V\\
HD 193281	&	20 20 28   &-29 11 50&		6.3	&	A2III\\
HD 207129	&	21 48 15	&-47 18 13	&	5.6	&	G0V\\
HD 216489	&	22 53 02 	&+16 50 28	&	5.9	&	K1.5II-IIIe\\
\enddata
\end{deluxetable}

\noindent{\bf References for the Supporting Online Material}

\noindent 1.  P. Kalas, D. Jewitt, {\it Astron. J.} {\bf 111}, 1347 (1996).\\

\noindent2. D. E. Backman, F. Paresce, in {\it Protostars and Protoplanets III},  
E. H. Levy and J. I. Lunine, Eds. (University of Arizona Press, Tucson, 1993), pp. 1253-1304.\\

\noindent3. M. C. Liu, B. C. Matthews, J. P. Williams, P. G. Kalas,  {\it Astrophys. J.}, in press (2004).\\

\noindent4. P. Kalas, D. Jewitt, {\it Astron. J.} {\bf 110}, 794 (1995). \\

\clearpage

\end{document}